\documentclass[aps,prl,showpacs,twocolumn,superscriptaddress,nofootinbib,floatfix,mamsmath,10pt]{revtex4}
\usepackage{epsfig,epsf,graphics,psfrag}
\usepackage{times}
\usepackage{color}
\usepackage{amsfonts,amssymb,stmaryrd,latexsym,amsmath}
\usepackage{hhline}

\newcommand{\be}{\begin{equation}}
\newcommand{\ee}{\end{equation}}
\newcommand{\ba}{\begin{eqnarray}}
\newcommand{\ea}{\end{eqnarray}}
\newcommand{\non}{\nonumber}

\newcommand{\al}{&\!\!\!}

\newcommand{\re}{\textrm{Re}}
\newcommand{\order}[2][M_\pi]{\mathcal{O}(#1^{#2})}

\newcommand{\Mo}[1]{ \overset{_\circ}{M}_{#1} }

\begin{document}
\title{Light Quark Mass Dependence in Heavy Quarkonium Physics}

\author{Feng-Kun Guo}
\email{fkguo@hiskp.uni-bonn.de} \affiliation{Helmholtz-Institut f\"ur Strahlen- und
             Kernphysik and Bethe Center for Theoretical Physics, \\
             Universit\"at Bonn,  D--53115 Bonn, Germany}

\author{Ulf-G. Mei{\ss}ner}
\email{meissner@hiskp.uni-bonn.de} \affiliation{Helmholtz-Institut f\"ur Strahlen-
und
             Kernphysik and Bethe Center for Theoretical Physics, \\
             Universit\"at Bonn,  D--53115 Bonn, Germany}
\affiliation{Institute for Advanced Simulation, Institut f\"{u}r Kernphysik
             and J\"ulich Center for Hadron Physics, \\
             Forschungszentrum J\"{u}lich, D--52425 J\"{u}lich, Germany}

\begin{abstract}
\noindent The issue of chiral extrapolations in heavy quarkonium systems is
discussed. We show that the light quark mass dependence of the properties of heavy
quarkonia  is not always suppressed. For quarkonia close to an open flavor
threshold, even a nonanalytic chiral extrapolation is needed. Both these nontrivial
facts are demonstrated to appear in the decay widths of the hindered $M1$
transitions between the first radially excited and ground state $P$-wave charmonia.
The results at a pion mass of about 500~MeV could deviate from the value at the
physical pion mass by a factor of two. Our findings show the necessity of
performing chiral extrapolations for  lattice simulations of heavy quarkonium
systems. Furthermore, lattice calculations of these transitions would also provide
a definite answer to the role of coupled-channel effects in heavy quarkonium
physics due to virtual heavy mesons.
\end{abstract}


\pacs{14.40.Pq, 12.38.Gc, 13.25.Gv}

\maketitle

Since the discovery of $J/\psi$, the physics of heavy quarkonium is an important
tool  for testing QCD. Because both the charm and bottom quark masses are much
larger than the nonperturbative scale $\Lambda_{\rm QCD}$, heavy quarkonia were
well described in the framework of potential models. However, in  recent years this
simple picture has been shattered, as quite a few charmonium states close to or
above the open charm thresholds were discovered, and many of their properties are
not expected from the potential models. For a recent comprehensive review, see
Ref.~\cite{Brambilla:2010cs}. The spectrum of heavy quarkonium has been intensively
studied using lattice simulations, using the quenched approximation in the early
stages and in full QCD in recent years; for example, see
Refs.~\cite{Gray:2005ur,Dowdall:2011wh,Burch:2009az,Meinel:2010pv,Mohler:2011ke,Daldrop:2011aa,Bali:2011rd,Liu:2012ze}.
While most of the calculations focus on the low-lying states, which we refer to as
the states below open heavy flavor thresholds, only a few calculations tackle the
problem of higher excited states~\cite{Dudek:2007wv,Bali:2011rd,Liu:2012ze}.
So far, all calculations of heavy quarkonium in full QCD are performed at light
quark masses larger than the physical values, or equivalently with unphysical pion
masses.
The simulations in Ref.~\cite{Liu:2012ze} are only performed at a single pion mass
$M_\pi=396$~MeV. Mixing of charmonia with pairs of open charm states are taken into
account in Ref.~\cite{Bali:2011rd} at three different pion masses ranging from
1~GeV down to 280~MeV, yet no chiral extrapolation to the physical pion mass was
performed. In addition to the spectrum, there have also been lattice simulations of
the charmonium~\cite{Dudek:2009kk,Chen:2011kp} and bottomonium radiative
transitions~\cite{Lewis:2011ti}. The quenched approximation is used in
Ref.~\cite{Dudek:2009kk}, and the calculations of Ref.~\cite{Chen:2011kp} were
performed at $M_\pi=485$~MeV.

Being bound states of a heavy quark and heavy antiquark, heavy quarkonia do not
contain any valence light quark. Thus, one would naively expect that the light
quark mass dependence of their properties would be suppressed, so that one can use
a simple linear formula in the light quark masses [remember for example that
$M_\pi^2 \propto (m_u+m_d)$ at leading order] for chiral extrapolation, as in
Refs.~\cite{Burch:2009az,Meinel:2010pv} for mass splittings. While this is true for
low-lying states, a similar simple extrapolation may not be reasonable for higher
excited states. The purpose of this Letter is to show that dramatic and even
nonanalytic dependences in the light quark masses can arise. Hence, for the excited
states that are close to open flavor thresholds, a formula that takes into account
the nonanalyticity should be utilized for chiral extrapolation. Furthermore, for
radiative transitions with strong coupled-channel effects, simulations at several
pion masses are necessary to extract the physical results.

The effects of light quarks in heavy quarkonium systems are due to quantum
fluctuations of the sea quarks. Sea quark and antiquark pairs are created and
annihilated in the color singlet heavy quarkonium. Low-energy fluctuations can be
described in the framework of chiral perturbation theory, which is the standard
tool for chiral extrapolations. The quarkonium states can be included as matter
fields. Let us focus on the quark mass dependence of the quarkonium mass.
\begin{figure}[tb]
\centering
\includegraphics[width=0.7\linewidth]{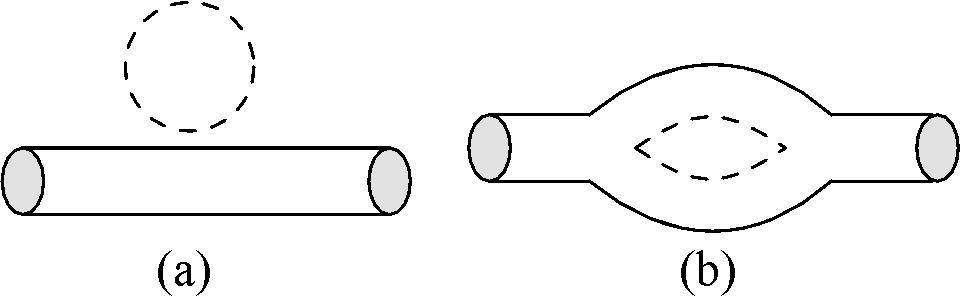}
\caption{Schematic diagrams of the creation and annihilation of sea quarks in a heavy quarkonium, with
solid and dashed lines representing heavy quarks and light sea quarks, respectively.
\label{fig:sea}}
\end{figure}
Two types of sea quark fluctuations are schematically depicted in
Fig.~\ref{fig:sea}; type (a) is disconnected and suppressed according to the
Okuba-Zweig-Iizuka rule; type (b) means that the heavy quark (antiquark) and the
virtual sea antiquark (quark) can form a color singlet state, a heavy meson
(antimeson); i.e. a virtual-heavy-meson--antimeson pair is created and annihilated
after a short propagation. There are certainly other contributions, such as the
doubly Okuba-Zweig-Iizuka suppressed processes that induce mixing of the heavy
quarkonium and a light meson. We expect that such contributions are less important,
and therefore do not consider them. Type (a) can be parameterized using an
effective chiral Lagrangian containing unknown low-energy constants. The resulting
quark mass dependence is analytic in the light quark masses up to chiral
logarithms, see Ref.~\cite{Grinstein:1996gm}. For instance, denoting the operator
annihilating a quarkonium field by $\psi$, a possible contribution would be
proportional to $\psi^\dag \psi\langle \chi_+ \rangle$ in the effective chiral
Lagrangian. Here, $\chi_+ = u^\dagger \chi u^\dagger + u\chi u$, $\langle~~\rangle$
is the flavor trace, and $\chi = 2B{\rm diag}\left(m_u,m_d,m_s\right)$ contains the
light quark mass matrix, where $B= |\langle 0 |\bar q q |0\rangle|/F^2$ and $F$ is
the pion decay constant in the chiral limit. The Goldstone boson field $\Phi$,
which contains the pions, eta and kaons in the SU(3) case, are included in
$u=\sqrt{U}$ and $U=\exp \left( i\sqrt{2}\Phi/F\right)$. At leading order
$\order{2}$, this term gives a contribution proportional to $B(m_u+m_d+m_s)$ to the
quarkonium mass, and at $\order{4}$, the chiral logarithm $M_\pi^4\log M_\pi^2$
will arise.

Complexity comes from type (b), which can lead to nonanalyticity as will be shown
below. Because the heavy quarkonium states are normally not far from the open
flavor thresholds, the open flavor mesons, at least the ground states, do not
necessarily decouple in a low-energy effective field theory (EFT) for heavy
quarkonium. In particular, the masses of many excited quarkonium states are very
close to the thresholds. 
In that case, one should consider coupled-channel effects due to coupling to the
open flavor mesons and antimesons in chiral extrapolation. As an example, let us
study $P$-wave charmonium states. They couple to the pseudoscalar and vector
charmed mesons in an $S$-wave with a coupling constant $g$. The self-energy due to
coupling to the charmed mesons with masses $m_1$ and $m_2$ is expressed in terms of
the scalar two-point loop function
\be%
\Sigma(P^2) \equiv \!\!\int\!\! \frac{d^4l}{(2\pi)^4}
\frac{i}{(l^2-m_1^2+i\epsilon) [(P-l)^2-m_2^2+i\epsilon]}.
\ee%
In the rest frame of the charmonium and taking the nonrelativistic approximation
for both propagators, we regularize the divergent loop with a three-momentum cutoff
$\lambda$
\be%
\Sigma(M^2,\lambda) =  \frac1{4\pi(m_1+m_2)} \left( -\frac{\lambda}{\pi} +
\frac12\sqrt{c-i\epsilon} \right),\label{eq:sigma0}
\ee%
where $M$ denotes the charmonium mass, $c = 2\mu_{12}b_{12}$ with
$\mu_{ij}=m_im_j/(m_i+m_j)$ the reduced mass and $b_{12} = m_1+m_2-M$. The mass
gets renormalized by the real part of the self-energy. Writing out the
$M_\pi$-dependence explicitly, we get
\be%
M(M_\pi) = M_0(\lambda,M_\pi) + g^2 m_1 m_2 \re \Sigma(M^2,\lambda,M_\pi),
\label{eq:renorm}
\ee%
where the factor $m_1m_2$ is required for correct normalization, and
\be%
M_0(\lambda,M_\pi) =  \Mo{0}(\lambda) + d(\lambda) M_\pi^2 + \order{4}
\label{eq:M0}
\ee%
is the bare mass. (Note that the mass shift due to the virtual loops is a
scale-dependent quantity, as can be seen in Eq.~(\ref{eq:renorm}), and not a
physical observable. For phenomenological studies of the charmed meson loops in the
charmonium spectrum, we refer to Refs.~\cite{Eichten:2005ga,Pennington:2007xr}.)
Both the chiral-limit bare mass $\Mo{0}(\lambda)$, and the coefficient $d(\lambda)$
depend on the cutoff, since the masses of open flavor heavy mesons $m_1$ and $m_2$
depend on the pion mass. For simplicity, we assume that the heavy mesons lie in the
same spin multiplet. Up to $\order{2}$, we have $m_{i} = \overset{_\circ}{m}_i +
h_1 {M_\pi^2}/{\overset{_\circ}{m}_i}$~\cite{Guo:2009ct}, where $h_1$ is a
dimensionless coefficient of order unity. Notice that
\ba%
\sqrt{c} = \sqrt{2\overset{_\circ}{\mu}_{12}\left( \delta +
\frac{h_1}{\overset{_\circ}{\mu}_{12}} M_\pi^2 \right) + \order{4}}, \label{eq:c}
\ea%
where $\overset{_\circ}{\mu}_{12} = \overset{_\circ}{m}_1
\overset{_\circ}{m}_2/(\overset{_\circ}{m}_1 + \overset{_\circ}{m}_2)$, and
$\delta=\overset{_\circ}{m}_1 + \overset{_\circ}{m}_2 - M$. Therefore, for the case
with $|\delta|\lesssim R\equiv M_\pi^2/\overset{_\circ}{\mu}_{12}$, the unitary cut
in the loop function Eq.~(\ref{eq:sigma0}) cannot be expanded in a polynomial in
$M_\pi$. Although $R\simeq20$~MeV is small at the physical pion mass, it is around
270~MeV for a pion mass of 500~MeV. As a result, there will be a cusp due to the
nonanalyticity at the point $M(M_\pi)=m_1(M_\pi)+m_2(M_\pi)$. Nonanalyticity due to
similar effects in the chiral extrapolation was discussed earlier for the
$\Delta$-resonance~\cite{Bernard:2009mw,Ledwig:2010ya} and the pion form
factor~\cite{Guo:2011gc}.

\begin{figure}[t]
\centering
\includegraphics[width=\linewidth]{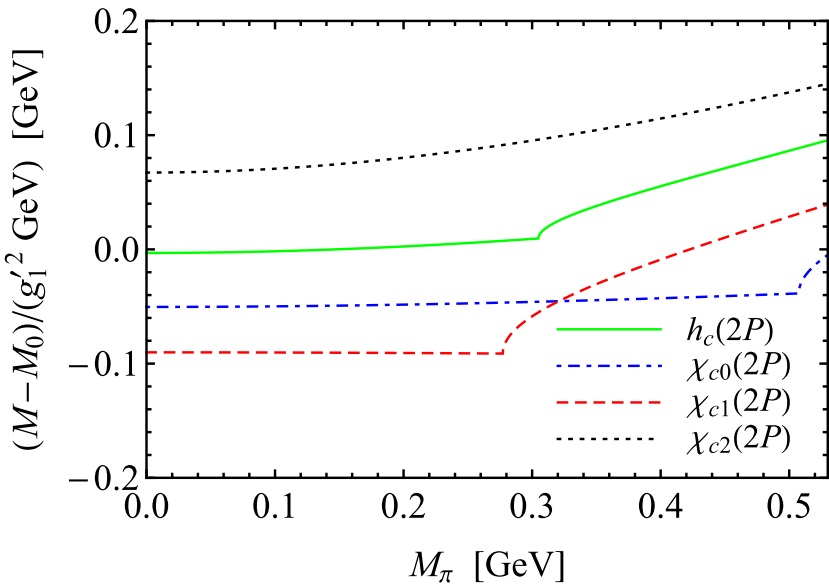}\\[2mm] 
\includegraphics[width=\linewidth]{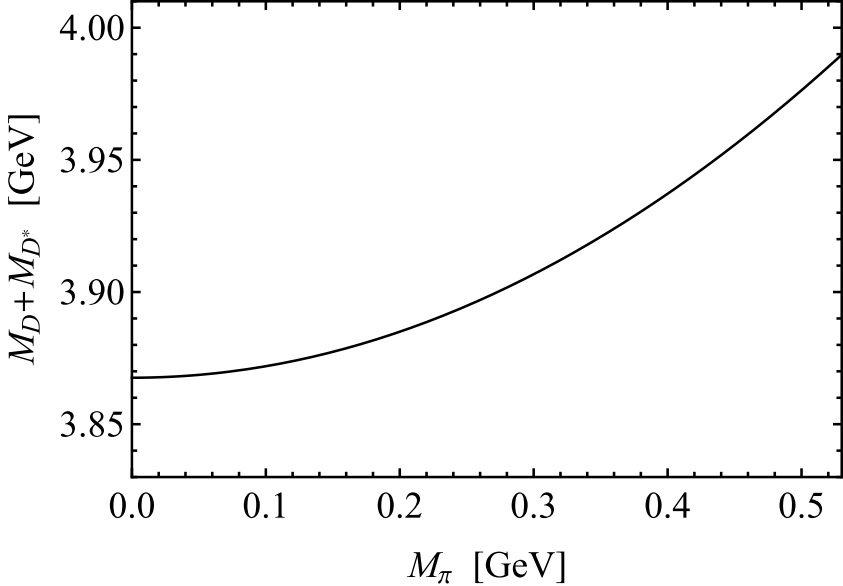}
\caption{Upper panel: Pion mass dependence of $M-M_0(\lambda)$ for the
$2P$ charmonia calculated with $\lambda=0.63$~GeV.
Lower panel: Pion mass dependence of the $D \bar D^*$ threshold.\label{fig:ms}}
\end{figure}

We use $h_1=0.44$ as determined from the SU(3) mass splittings of both the
pseudoscalar and vector charmed mesons~\cite{Hofmann:2003je,Guo:2008gp}. The
$M_\pi$-dependence of $M-M_0(\lambda)$ for the first radially excited $P$-wave
charmonia is plotted in Fig.~\ref{fig:ms}~(upper), where $\lambda=0.63$~GeV,
corresponding to $M-M_0(\lambda)=0$ for the $h_c'$ at the physical pion mass, is
used;
$g_1'$ is the coupling of the $2P$ charmonia to the charmed mesons, as defined
in~\cite{Guo:2010ak}, and its dimension is mass$^{-1/2}$. Using model values for
the masses of $\chi_{c0}'$, $\chi_{c1}'$ and $h_c'$ at the physical pion mass from
Ref.~\cite{Li:2009zu}, these states are above the coupled thresholds at the
physical pion mass. Increasing the pion mass, the charmed meson masses increase,
too. One expects the charmonium mass to increase more slowly than the charmed meson
thresholds. Therefore, for a charmonium with a mass slightly higher than the open
charm threshold at the physical pion mass, the charmonium mass should coincide with
the threshold at some larger pion mass. After that, the open charm mesons cannot go
on shell, and a cusp shows up because of the end of the unitary cut, as seen in
Fig.~\ref{fig:ms}~(upper); $\chi_{c2}'$ is always below the $D^*\bar D^*$
threshold, so that there is no cusp in the curve for this state.

For an $S$-wave charmonium, the nonanalyticity due to coupling to the pseudoscalar
and vector charmed mesons is less important, and even invisible. This is because
the coupling is in a $P$-wave. Of course, it can couple to a ground state charmed
meson and an orbitally excited state in an $S$-wave, as considered in
Ref.~\cite{Bali:2011rd}. However, the thresholds are far from the masses of the
$1S$ and $2S$ charmonia. In this case, the square root in Eq.~(\ref{eq:c}) can be
expanded in a polynomial in $M_\pi^2$. Hence, the $M_\pi$-dependence of the
quarkonium mass is given by Eq.~(\ref{eq:M0}) with redefined $\Mo{0}(\lambda)$ and
$d(\lambda)$.

It is instructive to briefly discuss possible hadronic molecules with a binding
energy much smaller than the pion mass. In this case, the bound state can be
described by an EFT with only contact terms analogous to that for the deuteron; for
example, see \cite{Hammer:2010kp}. Then the pion mass dependence of the mass of the
hadronic molecule is dominated by that of the masses of the constituents, as argued
in Ref.~\cite{Cleven:2010aw}. So, in the pion mass range where $X(3872)$ is a
$D\bar D^{*}$ bound state~\cite{Tornqvist:2004qy}, the $M_\pi$-dependence of its
mass should be approximately given by that of the threshold, as depicted in
Fig.~\ref{fig:ms}~(lower) at $\order{2}$. We will not calculate the deviation from
the threshold due to a small but finite binding energy here, but only point out
that a loosely bound state can easily become unbound by varying the interaction
strength. It is worth notice that the coupling constant $g$ in
Eq.~(\ref{eq:renorm}), which controls the strength of the cusp in the
$M_\pi$-dependence of the charmonium mass, is also a measure of the hadronic
molecular content of a given state~\cite{molecule1,molecule2}. However, it is
obvious that a quantitative treatment of the quark mass dependence of a hadronic
molecule requires a more refined approach than that given here. For example, see
Refs.~\cite{AlFiky:2005jd,Fleming:2007rp}.

The chiral corrections to the heavy quarkonium mass are always small compared to
the mass in the chiral limit. More noticeable is that there exist quantities in
heavy quarkonium physics whose pion mass dependence is strong. For these
quantities, chiral extrapolation is mandatory. One can imagine that in the mass
splittings between two heavy quarkonium states, the great bulk of the chiral-limit
masses cancels, and the chiral corrections are potentially large. However, as seen
in Eq.~(\ref{eq:M0}), it is not possible to give a parameter-free prediction for
its pion mass dependence even at $\order{2}$, since $d(\lambda)$ is scale
dependent. A prediction can only be made after fitting the parameters to
sufficiently large data. But there are indeed quantities whose quark mass
dependence is strong and can be predicted parameter free. A good example is given
by the decay widths of the hindered $M1$ transitions between the $2P$ and  $1P$
charmonium states. These transitions are shown to be dominated by coupled-channel
effects~\cite{Guo:2011dv} based on a nonrelativistic effective field theory
(NREFT)~\cite{Guo:2009wr,Guo:2010zk,Guo:2010ak}.

\begin{figure}[tb]
\centering
\includegraphics[width=0.7\linewidth]{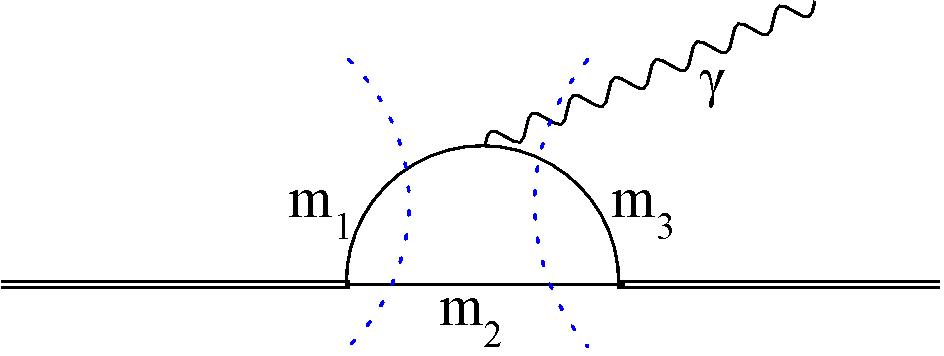}
\caption{Hadronic loop diagram. Double, solid and wiggly lines denote the
charmonia, charmed mesons, and photon, respectively. The dashed curves represent
the unitary cuts.
\label{fig:cuts}}
\end{figure}

As shown in Ref.~\cite{Guo:2011dv}, these transitions are dominated by triangle
diagrams at the hadronic level, with three intermediate charmed or anticharmed
mesons (see Fig.~\ref{fig:cuts}).
The decay amplitude is proportional to the three-point scalar loop function. It is
convergent, and the nonrelativistic expression reads
as~\cite{Guo:2010ak,Cleven:2011gp}
\ba%
I(q) \al\equiv\al i\int\!\frac{d^4l}{(2\pi)^4} 
\frac{1}{\left(l^2-m_1^2+i\epsilon\right) \left[(P-l)^2-m_2^2+i\epsilon\right]}
\non\\
\al\al\times \frac1{\left[(l-q)^2-m_3^2+i\epsilon\right] } \non\\
\al =\al {\cal N} \frac{1}{\sqrt{a}} \left[
\arctan\left(\frac{c'-c}{2\sqrt{a(c-i\epsilon)}}\right) \right.\non\\
\al \al \left. + \arctan\left(\frac{2a+c-c'}{2\sqrt{a(c'-a-i\epsilon)}}\right)
\right],
\ea%
where $P$ and $q$ are the momenta of the initial particle and the photon,
respectively, $b_{23}=m_2+m_3+q^0-M$ with $M$ the mass of the initial particle,
${\cal N}=\mu_{12}\mu_{23}/(16\pi m_1m_2m_3)$,
$$
a = \left(\frac{\mu_{23}}{m_3}\right)^2 \vec{ q}\,^2, \quad  c'=2\mu_{23}b_{23}+\frac{\mu_{23}}{m_3}\vec{ q}\,^2,
$$
and $c$ is as defined below Eq.~(\ref{eq:sigma0}). For small $a$, one may expand
the loop function out
\be%
I(q) = {\cal N} \frac{2}{\sqrt{c'}+\sqrt{c}} \left[1 +  {\cal
O}\left(\frac{a}{\min(c',(c'-c)/2)}\right) \right].
\ee%
It is then clear that two unitary cuts (see Fig.~\ref{fig:cuts}) lead to the main
contribution of the three-point loop. Corresponding to the two cuts, one may define
two velocities of the intermediate mesons. The velocity $v$ used in the NREFT power
counting~\cite{Guo:2010ak} should be understood as the average of these two
velocities. 
Following the discussion around Eq.~(\ref{eq:c}), we can expect cusps in the
$M_\pi$-dependence of the decay widths when the mass of the decaying particle
coincides with the coupled threshold.
\begin{figure}[t]
\centering
\includegraphics[width=\linewidth]{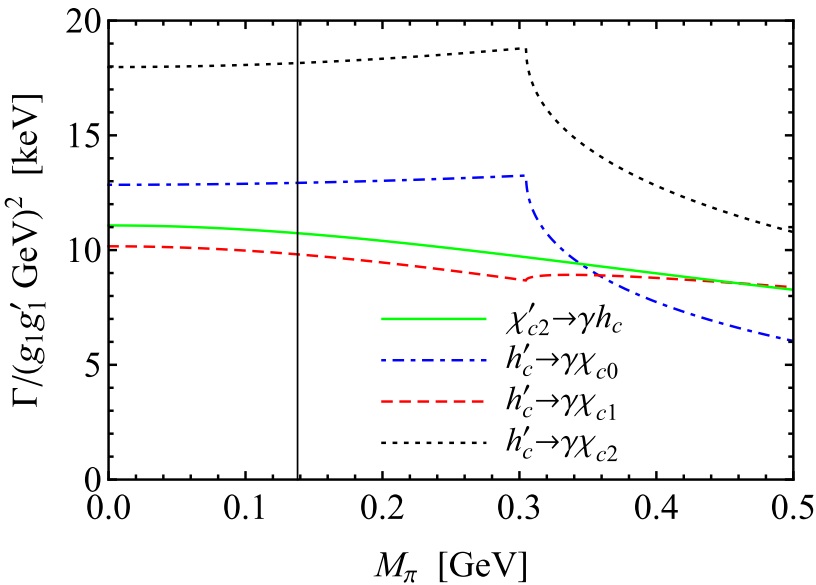}
\caption{Dependence of the widths of various hindered $M1$ transitions on the pion mass.
The vertical line denotes the physical pion mass.\label{fig:mpidep}}
\end{figure}
The results for $\Gamma(\chi_{c2}'\to h_c\gamma)$ and $\Gamma(h_c'\to
\chi_{cJ}\gamma)$ are shown in Fig.~\ref{fig:mpidep}, where we have neglected the
$M_\pi$-dependence of the charmonium masses and used the same model value 3908~MeV
for $M_{h_c'}$~\cite{Li:2009zu} as before.

The cusp in $\Gamma(h_c'\to \chi_{cJ}\gamma)$ appears at about $M_\pi=300$~MeV, the
same value as in Fig.~\ref{fig:ms} for $h_c'$. From Fig.~\ref{fig:mpidep}, one
finds a strong dependence on the pion mass. The value of $\Gamma(h_c'\to
\chi_{c0}\gamma)$ at a pion mass of 485~MeV, the same value as that used in the
lattice simulations for the $M1$ transitions between $S$-wave
charmonia~\cite{Chen:2011kp}, is only about half of that at the physical pion mass.
This observation highlights the necessity of chiral extrapolation of lattice
simulations for radiative transitions of heavy quarkonia and the necessity of small
pion masses in the simulations. Otherwise, the uncertainty due to unphysical pion
mass could be very large. Although parameter-free predictions for $M1$ transitions
between the $S$-wave heavy quarkonia are not possible, as noted in
Ref.~\cite{Guo:2011dv}, the charmed meson loops are still expected to be
crucial~\cite{Li:2007xr,Mehen:2011tp}. Thus, the pion mass dependence induced by
the virtual charmed mesons could introduce a large uncertainty due to the large
pion mass used in lattice simulations.

\begin{figure}[t]
\centering
\includegraphics[width=\linewidth]{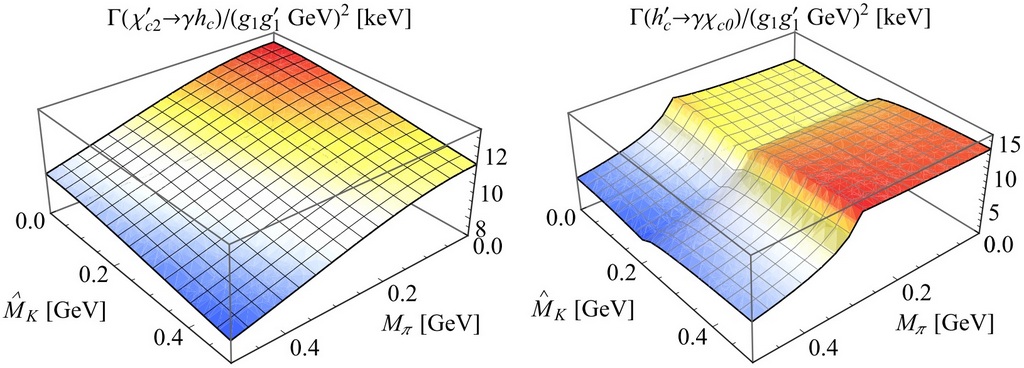}
\caption{ $M_\pi$ and $\hat M_K$ dependence of $\Gamma(\chi_{c2}'\to\gamma h_c)$
(left) and $\Gamma(h_c'\to\gamma\chi_{c0})$ (right).
\label{fig:3D}}
\end{figure}

One can further study the strange quark mass dependence, which translates into the
dependence on $\hat{M}_K^2=Bm_s$, where we use the same notation as
Ref.~\cite{Frink:2004ic}. As an example, we plot the simultaneous dependence on
$M_\pi$ and $\hat{M}_K$ of $\Gamma(\chi_{c2}'\to\gamma h_c)$ and
$\Gamma(h_c'\to\gamma\chi_{c0})$ in Fig.~\ref{fig:3D}, where we used $M_{D_s^{(*)}}
= \Mo{D^{(*)}} + 2 h_1 {\hat{M}_K^2}/{\Mo{D^{(*)}}}+\order[\hat{M}_K]{4}$. One
clearly sees the nonanalyticity in both the $M_\pi$ and $\hat{M}_K$ dependence of
the latter.

In conclusion, we have discussed  chiral extrapolations in heavy quarkonium
physics, especially for the higher excited states. These states are close to open
flavor thresholds. As a result, chiral extrapolation may be nonanalytic. This
observation is true for any excited hadron whose mass is in the neighborhood of an
$S$-wave-coupled hadronic threshold. For such a state, we propose to perform the
chiral extrapolation of the mass using
\be%
M(M_\pi) = \Mo{} + d M_\pi^2 + \sqrt{e+f M_\pi^2},
\ee%
where $\Mo{}$, $d$, $e$, and $f$ are parameters to be fit to the lattice data. For
states far away from any open flavor threshold, $e$ will be much larger than $f
M_\pi^2$ so that the square root can be expanded, and one may use only the first
two terms in the above equation up to $\order{2}$. Furthermore, we find that light
quark mass dependence is not always suppressed for heavy quarkonium systems. As an
example, we show that lattice results for the decay widths of the hindered $M1$
transitions between $P$-wave charmonia at a pion mass around 500~MeV can deviate by
a factor of 2 from the actual values at the physical pion mass. Simulations of
these transitions would also provide a nice test of the NREFT, and would be very
useful in identifying the coupled-channel effects, which might be the key to
understanding some long-standing puzzles in heavy quarkonium systems. If the
resulting pion mass dependences follow our predictions, they would also allow for
extraction of the product of coupling constants $g_1 g_1'$, which cannot be
measured directly.

\subsection*{Acknowledgments}
We want to thank Hans-Werner Hammer and Christoph Hanhart for valuable discussions.
This work is supported in part by the DFG and the NSFC through funds provided to
the sino-germen CRC 110 ``Symmetries and the Emergence of Structure in QCD'' and
the EU I3HP ``Study of Strongly Interacting Matter'' under the Seventh Framework
Program of the EU. U.-G. M. also thanks the BMBF for support (Grant No. 06BN9006).
F.-K. G. acknowledges partial support from the NSFC (Grant No. 11165005).


\begin{thebibliography}{99}


\bibitem{Brambilla:2010cs}
  N.~Brambilla {\it et al.},
  Eur.\ Phys.\ J.\ C {\bf 71}, 1534 (2011).


\bibitem{Gray:2005ur}
  A.~Gray 
  {\it et al.} [HPQCD and UKQCD Collaborations],
  Phys.\ Rev.\ D {\bf 72}, 094507 (2005).

\bibitem{Dowdall:2011wh}
  R.~J.~Dowdall {\it et al.}  [HPQCD Collaboration],
  Phys.\ Rev.\ D {\bf 85}, 054509 (2012).

\bibitem{Burch:2009az}
  T.~Burch 
  {\it et al.} [Fermilab Lattice and MILC Collaborations],
  Phys.\ Rev.\ D {\bf 81}, 034508 (2010).

\bibitem{Meinel:2010pv}
  S.~Meinel,
  Phys.\ Rev.\ D {\bf 82}, 114502 (2010).

\bibitem{Mohler:2011ke}
  D.~Mohler and R.~M.~Woloshyn,
  Phys.\ Rev.\ D {\bf 84}, 054505 (2011).

\bibitem{Daldrop:2011aa}
  J.~O.~Daldrop, C.~T.~H.~Davies and R.~J.~Dowdall,
  Phys.\ Rev.\ Lett.\  {\bf 108}, 102003 (2012).


\bibitem{Bali:2011rd}
  G.~S.~Bali, S.~Collins and C.~Ehmann,
  Phys.\ Rev.\ D {\bf 84}, 094506 (2011).


\bibitem{Liu:2012ze}
  L.~Liu 
  {\it et al.} [Hadron Spectrum Collaboration],
  arXiv:1204.5425 [hep-ph].


\bibitem{Dudek:2007wv}
  J.~J.~Dudek, R.~G.~Edwards, N.~Mathur and D.~G.~Richards,
  Phys.\ Rev.\ D {\bf 77}, 034501 (2008).




\bibitem{Dudek:2009kk}
  J.~J.~Dudek, R.~Edwards, C.~E.~Thomas,
  Phys.\ Rev.\ D {\bf 79}, 094504 (2009).

\bibitem{Chen:2011kp}
  Y.~Chen  {\it et al.},
  Phys.\ Rev.\ D {\bf 84}, 034503 (2011).

\bibitem{Lewis:2011ti}
  R.~Lewis, R.~M.~Woloshyn,
  Phys.\ Rev.\ D {\bf 84}, 094501 (2011).


\bibitem{Grinstein:1996gm}
  B.~Grinstein and I.~Z.~Rothstein,
  Phys.\ Lett.\ B {\bf 385}, 265 (1996).


\bibitem{Eichten:2005ga}
  E.~J.~Eichten, K.~Lane, C.~Quigg,
  Phys.\ Rev.\ D {\bf 73}, 014014 (2006).

\bibitem{Pennington:2007xr}
  M.~R.~Pennington, D.~J.~Wilson,
  Phys.\ Rev.\ D {\bf76}, 077502 (2007);
  T.~Barnes, E.~S.~Swanson,
  Phys.\ Rev.\ C {\bf77}, 055206 (2008);
  B.-Q.~Li, C.~Meng, K.-T.~Chao,
  Phys.\ Rev.\ D {\bf80}, 014012 (2009).


\bibitem{Guo:2009ct}
  F.-K.~Guo, C.~Hanhart and U.-G.~Mei{\ss}ner,
  Eur.\ Phys.\ J.\ A {\bf 40}, 171 (2009).





\bibitem{Bernard:2009mw}
  V.~Bernard, D.~Hoja, U.-G.~Mei{\ss}ner and A.~Rusetsky,
  JHEP {\bf 0906}, 061 (2009).

\bibitem{Ledwig:2010ya}
  T.~Ledwig, V.~Pascalutsa and M.~Vanderhaeghen,
  Phys.\ Rev.\  D {\bf 82}, 091301 (2010).

\bibitem{Guo:2011gc}
  F.-K.~Guo, C.~Hanhart, F.~J.~Llanes-Estrada and U.-G.~Mei{\ss}ner,
  Phys.\ Lett.\  B {\bf 703}, 510 (2011).

\bibitem{Guo:2008gp}
  F.-K.~Guo, C.~Hanhart, S.~Krewald and U.-G.~Mei{\ss}ner,
  Phys.\ Lett.\ B {\bf 666}, 251 (2008).

\bibitem{Hofmann:2003je}
  J.~Hofmann and M.~F.~M.~Lutz,
  Nucl.\ Phys.\ A {\bf 733}, 142 (2004).


\bibitem{Guo:2010ak}
  F.-K.~Guo, C.~Hanhart, G.~Li, U.-G.~Mei\ss ner, Q.~Zhao,
  Phys.\ Rev.\ D {\bf 83}, 034013 (2011).


\bibitem{Li:2009zu}
  B.~Q.~Li and K.~T.~Chao,
  Phys.\ Rev.\  D {\bf 79}, 094004 (2009).

\bibitem{Hammer:2010kp}
  H.-W.~Hammer and L.~Platter,
  Ann.\ Rev.\ Nucl.\ Part.\ Sci.\  {\bf 60}, 207 (2010).

\bibitem{Cleven:2010aw}
  M.~Cleven, F.-K.~Guo, C.~Hanhart and U.-G.~Mei{\ss}ner,
  Eur.\ Phys.\ J.\  A {\bf 47}, 19 (2011).

\bibitem{Tornqvist:2004qy}
  N.~A.~T\"ornqvist,
  Phys.\ Lett.\ B {\bf 590}, 209 (2004).

\bibitem{molecule1}
  S.~Weinberg,
  Phys.\ Rev.\  {\bf 130}, 776 (1963);
  Phys.\ Rev.\  {\bf 131}, 440 (1963);
  Phys.\ Rev.\  {\bf 137}, B672 (1965).

\bibitem{molecule2}
  V.~Baru, J.~Haidenbauer, C.~Hanhart, Yu.~Kalashnikova and A.~E.~Kudryavtsev,
  Phys.\ Lett.\  B {\bf 586}, 53 (2004).


\bibitem{AlFiky:2005jd}
  M.~T.~AlFiky, F.~Gabbiani and A.~A.~Petrov,
  Phys.\ Lett.\ B {\bf 640}, 238 (2006).

\bibitem{Fleming:2007rp}
  S.~Fleming, M.~Kusunoki, T.~Mehen and U.~van Kolck,
  Phys.\ Rev.\ D {\bf 76}, 034006 (2007).


\bibitem{Guo:2011dv}
  F.-K.~Guo and U.-G.~Mei{\ss}ner,
  Phys.\ Rev.\ Lett.\  {\bf 108}, 112002 (2012).



\bibitem{Guo:2009wr}
  F.-K.~Guo, C.~Hanhart and U.-G.~Mei{\ss}ner,
  Phys.\ Rev.\ Lett.\  {\bf 103}, 082003 (2009)
  [Erratum, {\it ibid} {\bf 104}, 109901 (2010)].

\bibitem{Guo:2010zk}
  F.-K.~Guo, C.~Hanhart, G.~Li, U.-G.~Mei{\ss}ner and Q.~Zhao,
  Phys.\ Rev.\  D {\bf 82}, 034025 (2010).

\bibitem{Cleven:2011gp}
  M.~Cleven, F.-K.~Guo, C.~Hanhart, U.-G.~Mei{\ss}ner,
  Eur.\ Phys.\ J.\ A {\bf 47}, 120 (2011).

\bibitem{Li:2007xr}
  G.~Li and Q.~Zhao,
  Phys.\ Lett.\ B {\bf 670}, 55 (2008);
  Phys.\ Rev.\ D {\bf 84}, 074005 (2011).

\bibitem{Mehen:2011tp}
  T.~Mehen and D.-L.~Yang,
  Phys.\ Rev.\ D {\bf 85}, 014002 (2012).


\bibitem{Frink:2004ic}
  M.~Frink and U.-G.~Mei{\ss}ner,
  JHEP {\bf 0407}, 028 (2004).

\end{thebibliography}
\end{document}